\documentclass[sn-mathphys-num]{sn-jnl}%

\usepackage{graphicx}%
\usepackage{multirow}%
\usepackage{amsmath,amssymb,amsfonts}%
\usepackage{amsthm}%
\usepackage{mathrsfs}%
\usepackage[title]{appendix}%
\usepackage{xcolor}%
\usepackage{textcomp}%
\usepackage{manyfoot}%
\usepackage{booktabs}%
\usepackage{algorithm}%
\usepackage{algorithmicx}%
\usepackage{algpseudocode}%
\usepackage{listings}%

\usepackage[utf8]{inputenc}
\usepackage[T1]{fontenc}
\usepackage{graphicx}
\usepackage{physics}
\usepackage[dvipsnames]{xcolor}
\usepackage{booktabs}
\usepackage{multirow}
\usepackage{bbm}
\usepackage{amsfonts}
\usepackage{amssymb}
\usepackage{hyperref}
\hypersetup{ 
    colorlinks=true,
    linkcolor=blue,
    breaklinks=true,
    citecolor=blue,
    filecolor=blue,
    urlcolor=blue}
\usepackage{url}

\newcommand{\blambda}{\boldsymbol{\lambda}}

\newcommand{\by}{\boldsymbol{y}}
\newcommand{\bkappa}{\boldsymbol{\kappa}}
\newcommand{\Graph}{{\mathcal{G}}}
\newcommand{\edges}{{\mathcal{E}}}
\newcommand{\vertices}{{\mathcal{V}}}

\begin{document}
\title[Analog QAOA with Bayesian Optimisation on a neutral atom QPU]{Analog QAOA with Bayesian Optimisation on a neutral atom QPU}

\author*[1,2]{\fnm{Simone Tibaldi} }\email{simonetibaldi2@unibo.it}

\author[3]{\fnm{Lucas} \sur{Leclerc}}\email{lucas.leclerc@pasqal.com}

\author[1,2]{\fnm{Davide} \sur{Vodola}}\email{davide.vodola@unibo.it}

\author[4]{\fnm{Edaordo} \sur{Tignone}}\email{edoardo.tignone@leitha.eu}

\author[1,2]{\fnm{Elisa} \sur{Ercolessi}}\email{elisa.ercolessi@unibo.it}

\affil[1]{\orgdiv{Dipartimento di Fisica e Astronomia ``Augusto Righi''}, \orgname{Universit\`a di Bologna}, \orgaddress{\street{Via Irnerio 48}, \city{Bologna}, \postcode{I-40127}, \country{Italy}}}

\affil[2]{\orgdiv{INFN}, \orgname{Sezione di Bologna}, \orgaddress{\street{Via Carlo Berti-Pichat 6/2}, \city{Bologna}, \postcode{I-40127}, \country{Italy}}}

\affil[3]{\orgdiv{Pasqal}, \orgname{Organization}, \orgaddress{\street{24 rue Emile Baudot}, \city{Palaiseau}, \postcode{91120 }, \country{France}}}

\affil[4]{\orgdiv{Leithà S.r.l.}, \orgname{Unipol Group}, \orgaddress{\city{Bologna}, \country{Italy}}}

\abstract{This study explores the implementation of the Quantum Approximate Optimisation Algorithm (QAOA) in its analog form using a neutral atom quantum processing unit to solve the Maximum Independent Set problem.  Our QAOA protocol leverages the natural encoding of problem Hamiltonians by Rydberg atom interactions, while employing Bayesian Optimisation  to navigate the quantum-classical parameter space effectively under the constraints of hardware noise and resource limitations. We evaluate the approach through a combination of numerical simulations and experimental runs on Pasqal's first commercial quantum processing unit, Orion Alpha, demonstrating effective parameter optimisation and noise mitigation strategies, such as selective bitstring discarding and detection error corrections. Results show that a limited number of measurements still allows for a quick convergence to a solution, making it a viable solution for resource-efficient scenarios.}


\maketitle

\section*{Introduction}

Quantum computers, currently in their noisy intermediate-scale quantum (NISQ) era \cite{preskill2018nisq}, hold the promise of addressing computational problems that are intractable for classical systems. However, the utility of quantum algorithms on NISQ devices is fundamentally constrained by the prevalence of noise, decoherence, and limited qubit connectivity. As a result, significant research efforts are focused on understanding the sources and dynamics of noise in quantum hardware \cite{knill2005quantum, shor1995scheme} and developing classical methods to mitigate its effects, either during execution or in post-processing stages \cite{kandala2019error, temme2017error,Ball2021-gd}.

Rydberg atom arrays, which naturally encode problem Hamiltonians into their physical interactions \cite{pagano2020quantum}, offer an attractive platform for solving graph-based optimisation problems \cite{Dalyac2024-qc,leclerc:tel-04745992}. Different approaches have been proposed to use hybrid protocols on this kind of platforms. In \cite{ebadi2022quantum}, the authors analyse the performance of several quantum variational algorithms, exploring a class of graphs with programmable connectivity, while in \cite{schuetz2025qredumisquantuminformedreductionalgorithm},  a quantum-informed reduction algorithm (qReduMIS, where the quantum computer is used as a co-processor for a classical reduction process) is benchmarked on some platforms commercially available online. 

Hybrid quantum-classical approaches leverage classical optimisation methods to improve quantum algorithm performance despite noisy conditions. Among these, the Quantum Approximate Optimisation Algorithm (QAOA) \cite{farhi2014quantum, Zhou20} stands out for its remarkable ability to solve combinatorial optimisation problems by iteratively optimising quantum parameters.  The QAOA protocol can be realized in an analog approach, via the control of global parameters of the problem Hamiltonian and, notably, its performance heavily depends on the quality of parameter tuning. Heuristic strategies have been proposed to increase the efficiency of the algorithm on near term devices, which include: interpolation and functional techniques to initialize the data at level $p+1$ starting from the results at level $p$  \cite{Zhou20}; the use of classical optimisation techniques like Bayesian Optimisation (BO) \cite{shahriari2016taking}, known for its efficiency in scenarios with costly function evaluations~\cite{Snoek2012} and which have been proved to be suited for hybrid quantum-classical algorithms~\cite{TibaldiBO2023, Tamiya2022, Cheng2024}.

This work investigates the use of QAOA on a Rydberg atom platform, focusing on the Maximum Independent Set (MIS) problem~\cite{DP1999}. We investigate the feasibility of implementing the QAOA algorithm in its analog implementation on a real commercial device, namely Pasqal's neutral atom quantum processing unit (QPU) \cite{Henriet2020,Browaeys2020}. The analog nature of this implementation introduces constraints on the control parameters,  that might result in a lower efficiency of the classical optimization protocol. Thus innovative strategies to adapt classical optimisation to the quantum system's specific limitations are requested. 

We chose to use BO, which is able to efficiently explore the parameter space while minimising quantum resource usage. Indeed, 
in \cite{TibaldiBO2023}, it has been shown that BO is an adaptive method that allows for a significant reduction in the number of calls to the quantum hardware, working well in 
the regime of slow circuit repetition rates and with a limited number of measurements, while being resilient to uncorrelated quantum noise in the specification of the parameters.

Additionally, we incorporate post-processing techniques to control the impact of detection noise, such as bitstring discarding and error mitigation~\cite{Cai2023}. These strategies, combined with numerical simulations and experimental validations, enable to evaluate the algorithm performance across varying problem sizes and levels of hardware noise.

By addressing the challenges of applying noise mitigation, efficiently optimising parameters while respecting hardware constraints, we show how to navigate the challenges of quantum optimisation in the NISQ era. 

The paper is structured as follows. In Section~\ref{sec:algo} we introduce the hybrid algorithm, explaining in details the mapping of the classical problem to atomic registers (\ref{ssec:MIS}), the analog implementation of QAOA (\ref{ssec:analogqaoa}) and the specific cost functions used to tackle MIS problems (\ref{ssec:score-MIS}). All together, these elements form a variational loop between a QPU and a classical bayesian optimiser (\ref{ssec:loop}). Section~\ref{sec:results} is dedicated to the results first obtained in numerical simulations (\ref{ssec:simul}) to study the scaling of the method and then benchmarked on Pasqal QPU, Orion Alpha (\ref{ssec:benchmark}) allowing us to infer post-processing  techniques to mitigate detection errors (\ref{ssec:errors}). Finally, we perform closed variational loops directly on the QPU (\ref{ssec:exp}) and show that our algorithm is able to identify the solution on small and larger graph instances with few optimisation steps and limited number of measurements. The three appendices provide a description of the techniques used in the various sections, providing a brief description of analogue computation with Rydberg atoms, the Bayesian optimisation method and correction of detection errors.

\begin{figure*}
    \centering
    \includegraphics[width=\linewidth]{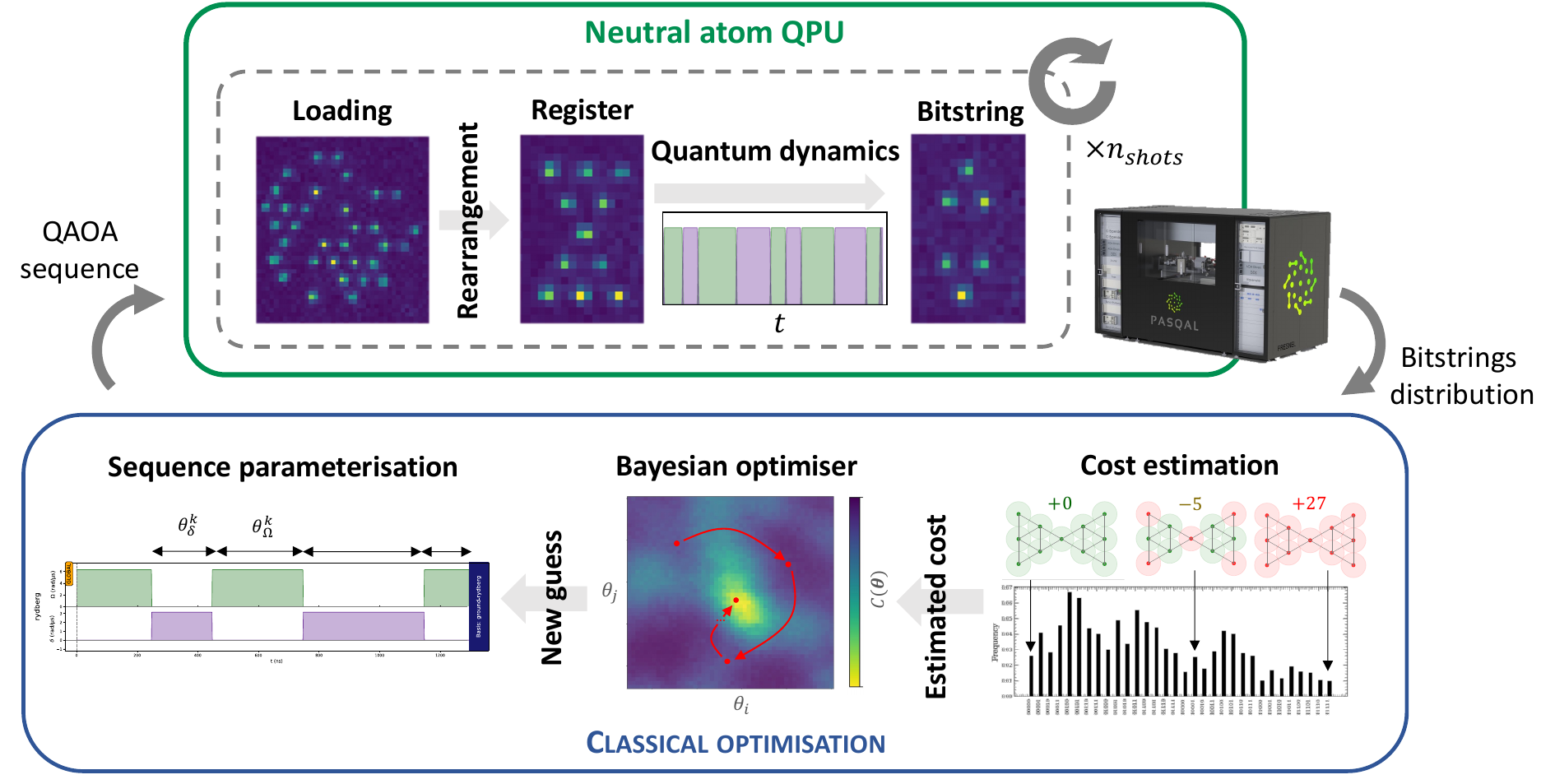}
    \caption{\textbf{QAOA variational loop on a neutral atom quantum processor}. The classical optimisation procedure, shown in the lower panel, is composed of: a cost estimation method, favouring MIS configurations; a decision-making Bayesian optimiser, navigating the parameter space with an exploitation-exploration strategy; and a sequence parametrisation step, building a new driving protocol from an initial  parameter instance, to be sent to a neutral atom QPU. In the latter, shown in the upper panel, atoms are loaded, arranged into positions to reproduce the graph considered, and the quantum system is let to evolve according to the built driving protocol. Measuring the system provides one bitstring and a distribution is acquired by repeating the quantum procedure several times before being sent to the classical part. This hybrid loop runs until the QAOA converges.}
    \label{fig:QAOA}
\end{figure*}

\section{Algorithm description}
\label{sec:algo}
In this section, we introduce the specifics of our implementation of QAOA that takes into consideration the use of Bayesian optimisation for the classical optimisation part and the analog computation available on a Rydberg atoms platform for the quantum part.

\subsection{Mapping MIS solutions to atom register}
\label{ssec:MIS}
Given a graph $\Graph=(\vertices, \edges)$, an independent set is defined as a subset $S$ of the vertices such that no two vertices of $S$ share an edge in $\Graph$. A MIS $S^*$ corresponds to an independent set of maximum cardinality. For a given set, we attribute a status $\boldsymbol{z}$ to a set of vertices, where $z_i = 1$ if the vertex $i$ belongs to the set, and $z_i=0$ otherwise. The configuration $S^*$ minimises the following cost function 
\begin{equation}
\label{eq:mis_hamiltonian}
    C_\Graph(\boldsymbol{z}) = -\sum_{i\in\vertices} z_i + c \sum_{(i,j) \in \edges} z_i z_j,
\end{equation}
where $c\gg 1$, as the last term penalises the selection of adjacent vertices in the set. In addition, $C_\Graph$ can be interpreted as an energy function where $z_i \in \{0,1\}$ is an Ising spin variable.

The binary perspective of the Rydberg blockade effect, solely depending on the threshold $r_b$ as described in App.\,\ref{app:Analog}, natively mimics the topology of a Unit-Disk (UD) graph, where two vertices of which share an edge if they lie within a threshold distance from each other in the Euclidean plane. Therefore, utilising the Hamiltonian in Eq.\,\ref{eq:ising-hamiltonian} as a resource, by either embedding a UD graph into an atomic register and conversely defining a graph from a spatial configuration of atoms and a fixed $r_b$, we can engineer the Hamiltonian
\begin{equation}
    \hat{H}_\delta/\hbar = \delta C_\Graph(\hat{\boldsymbol{n}})=-\delta \sum_{i\in \vertices} \hat{n}_i + U\sum_{(i,j) \in \edges} \hat{n}_i \hat{n}_j,
\end{equation}
with $U=c\delta$. By construction, the ground-state of the Hamiltonian $H_\delta$ encodes the MIS of $\Graph$ when $\delta>0$. Therefore, preparing and sampling such a solution state enables to retrieve \textit{bitstrings} $\boldsymbol{z}$ minimising $C_\Graph$ and thus solving the MIS problem for $\Graph$.

In this work, we reproduce given unweighted UD graphs with atoms, i.e. each vertex is represented by an atom and all adjacent pairs of vertices are separated by the same distance, in order to experience the same Rydberg blockade. A convenient embedding can be achieved using a regular triangular lattice trap layout with spacing $a$. Reproducing a graph only requires selecting which traps of the triangular layout need to be filled. This also enables to have a homogeneous $U=C_6/\hbar/a^6$ over adjacent pairs and justify the Rydberg blockade approximation as next nearest neighbouring pairs interacts with strength $U/\sqrt{3}^6$.  

\subsection{Analog QAOA with Rydberg atoms}
\label{ssec:analogqaoa}

QAOA is a well-known algorithm in the quantum optimisation paradigm. It consists of $p$ successive layers of unitary evolution, each layer being governed by two distinct Hamiltonians, called a cost $\hat C$ and a mixing $\hat M$ one, with variable angles $\theta^{2i-1}$ and $\theta^{2i}$ respectively ($i=1,2, \cdots, p$). We will denote with  $\boldsymbol{\theta} = (\theta^1, \theta^2, \cdots, \theta^{2p})$ the total set of parameters.  The algorithm initialises by preparing the system in $\ket{\psi_0} = \ket{+}^{\otimes N}=2^{-N/2} \sum_{\boldsymbol{z}\in\mathbb{B}^N} \ket{\boldsymbol{z}}$, representing the ground state of the mixing Hamiltonian, often chosen as $\hat{M} = \sum_i \hat{\sigma}_i^x$. By evolving from this state to $\ket{\psi(\boldsymbol{\theta})}$ following Eq.\,\ref{eq:evolution}, and with the aid of a classical optimiser, it becomes possible {\color{red} \cite{farhi2014quantum}}, as $p \rightarrow \infty$, to determine an optimal set of angles $\boldsymbol{\theta}^{*}$ such that:
    \begin{equation}
        |\psi(\boldsymbol{\theta}^{*})\rangle = \text{argmin}_{\psi}\bra{\psi}\hat{C}\ket{\psi}
    \end{equation}

In the Rydberg atoms version, the cost operator $\hat C=C_\mathcal{G}(\hat{\boldsymbol{n}})$ can only be fully replicated for specific graphs and the mixing part can be tackled with resonant pulses of amplitude $\Omega$. A major limitation preventing a straightforward implementation of QAOA on a Rydberg setup lies in the inability to turn off the interaction component of the Rydberg Hamiltonian during the mixing phase of each layer. This also implies that a clean preparation of $\ket{\psi_0}$ might require a more complex protocol or be even not possible. Applying pulses with $\Omega\gg U$ enables to neglect the interaction effects but the maximum amplitude reachable on current devices is only around $\Omega_{\rm max}/2\pi\sim$ few MHz. Consequently, this requirement mandates low values of $U$ and hence of $\delta$, which in turn prolongs the sequence and renders the evolution vulnerable to decoherence. One can nonetheless program an evolution with a QAOA-like protocol, applying series of resonant pulses with fixed $\Omega$, producing 
\begin{equation}
    \hat H_\Omega/\hbar=\frac{\Omega}{2}\hat M + U\sum_{i<j} \hat{n}_i \hat{n}_j,
\end{equation}
interleaved by periods of free evolution under fixed detuning $\delta$, producing $\hat H_\delta$, similar to approaches used in analog quantum annealers~\cite{Barraza2022}. Each layer durations, $\theta_\Omega^k/\Omega$ and $\theta_\delta^k/\delta$, constitute $2$ parameters to optimise on, amounting to a vector of optimisation parameters $\boldsymbol{\theta}$ with length $|\boldsymbol{\theta}|=2p$. 

The durations of each layer, $\theta_\Omega^k/\Omega$ and $\theta_\delta^k/\delta$, constitute two parameters to optimise. Furthermore, several hardware constraints bound the values that $\boldsymbol{\theta}$ can take. The rise time of the pulse-shaping device limits the minimum pulse duration to $100~\text{ns}$. To mitigate performance losses due to the various decoherence effects discussed in Appendix~\ref{app:noise}, the total duration of a pulse sequence is restricted to $4~\mu\text{s}$. This requires imposing linear constraints between the various layers at larger depths. Additionally, the maximum duration of a single pulse is set to $2\pi/\Omega$, which for $\Omega/2\pi = 1~\text{MHz}$ corresponds to a bound of $1~\mu\text{s}$. Therefore, we restrict the study to QAOA protocol with depths lower than $5$, thus using no more than $10$ parameters, with $\boldsymbol{\theta}$ further constrained by the aforementioned linear conditions.

The starting state $\ket{\psi_0}$ is obtained by applying an initial mixing pulse with $\theta^0_\Omega=\pi/2$. This is similar, but not exactly equivalent due to the always on interaction terms, to initialising all the qubits in $\ket{+}$. The quantum dynamics then produces the final state:
\begin{equation}
    \ket{\psi(\boldsymbol{\theta})}=\prod_{k=1}^p \exp[-i\theta_\Omega^k \hat{H}_\Omega/\hbar]\exp[-i\theta_\delta^k \hat{H}_\delta/\hbar] \ket{\psi_0}
    \label{eq:qaoa-dynamics}
\end{equation} 
which is sampled $n_{\rm shots}$ times to estimate the various MIS-related costs described in the next section. 

\subsection{Characterising the closeness to MIS}
\label{ssec:score-MIS}
In order to characterise how close a given quantum state $\ket{\psi(\boldsymbol{\theta})}$ (or associated probability distribution) stands from the solution state $\ket{S^*}$ (or from the manifold $\mathcal{H}_{S^*}$ of solution states if the MIS is degenerated), several figures of merit can be introduced: 
\begin{itemize}
    \item The \textit{fidelity} $F=\sum_{i\in\mathcal{H}_{S^*}}|\langle i|\psi(\boldsymbol{\theta})\rangle|^2$. 
    \item The \textit{approximation ratio} 
    \begin{equation}
        R={\color{red}-}E(\boldsymbol{\theta})/|S^*| \equiv\expval{C_\Graph(\hat{\boldsymbol{n}})}{\psi(\boldsymbol{\theta})}/|S^*|,
    \end{equation} with $E(\boldsymbol{\theta})$ the energy of the quantum system and $R$ can be normalised when knowing beforehand the MIS size.
    \item When the algorithm is able to measure the solution state $\ket{S^*}$ as the most probable, we define the \textit{solution ratio} $S_r$ as
    \begin{equation}
    \label{eq:solutionRatio}
        S_r  = \frac{p(S^*)}{p(S^{2^{nd}})}
    \end{equation} 
where $p(S^*)$ is the probability of measuring the solution state $\ket{S^*}$, while $p(S^{2^{nd}})$ the measured probability of the second most likely state. Thus, $S_r\geq 1$ quantifies how much the solution stands out from the other states.  We set $S_r=0$ otherwise, to indicate that the solution was not the most measured state.
\end{itemize}
We will use these metrics to assess the performance of the algorithm presented in the following section. 
We stress that the solution ratio $S_r$
  is intended here as a measure of how strongly the maximum independent set is isolated from competing states in the output distribution. Unlike hardness parameters introduced in Refs.\cite{Finzgar2024,Andrist_2023}, which are designed to normalize performance across heterogeneous problem instances, our goal is not to compare instances of different intrinsic difficulty but rather to quantify the sharpness of the solution peak for the specific graphs considered. For our graph family, the hardness parameter typically takes uniformly low values, making it less informative for this purpose.

\begin{figure*}[t]
    \centering
    \includegraphics[width = \linewidth]{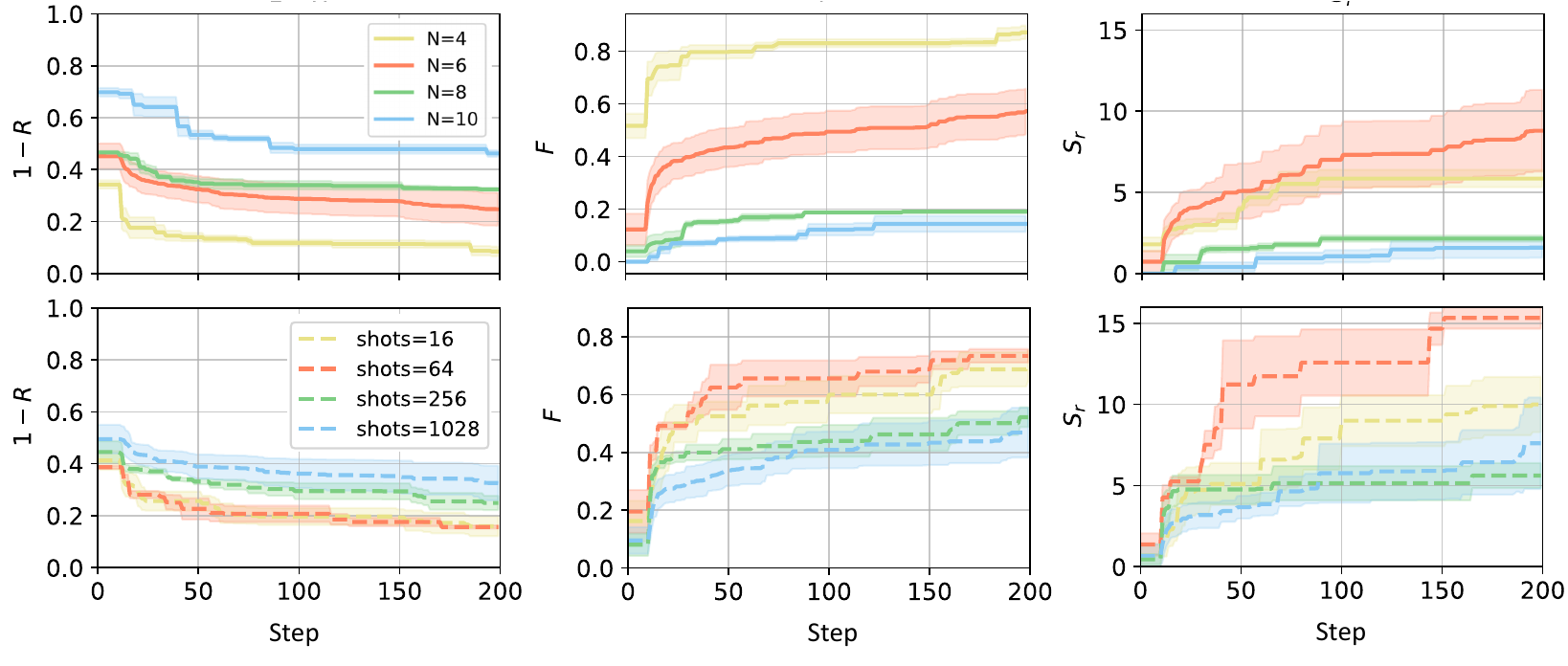}
    \caption{\textbf{QAOA loops numerically simulated using \texttt{Pulser}}. The evolution of the averaged approximation ratio $1-R$, the fidelity $F$ and the solution ratio $S_r$ (explained in the text) during optimisation are shown for various problem instances. In the first row, continuous lines show the results, varying the graph size with $N=4, 6, 8, 10$, while keeping fixed the number of shots to 1028. In the second row, dashed lines represent the results for a fixed graph of size, $N=6$, and increasing number of shots: $16, 64, 256, 1028$. The values shown are obtained averaging over 10 runs, light colour areas show the results $\pm \frac{1}{2}$ of standard deviation. Each run starts with $M=10$ training points and continues for $n_{\rm steps} = 190$ steps.}
    \label{fig:SimNqubits}
\end{figure*}

\subsection{Variational loop of QAOA}
\label{ssec:loop}
The variational loop of the analog version of QAOA is showcased in Fig.\,\ref{fig:QAOA}. The algorithm is essentially a hybrid loop between an optimisation procedure fully handled by classical resources (CPU) and a QPU able to perform the quantum dynamics of Eq.\,\ref{eq:qaoa-dynamics}. Communication between the two entities boils down to new parameter instances being sent from the CPU to the QPU, and to measurements being sent back by the QPU and used by the CPU to provide a better set of parameters. The loop is iterated until a convergence criterion is met, the simplest being until a fixed number of iterations $n_{\rm steps}$ has been performed.

The cost estimation procedure can be seen as a noisy and expensive-to-evaluate black-box function. The Bayesian optimisation method \cite{brochu2010tutorial} outlined in App.\,\ref{app:Bayes} proves particularly advantageous in this specific scenario, thanks to its resilience against noisy evaluations, its efficiency even within a limited budget of iterations, and its gradient-free approach.  Similar Bayesian optimization strategies have been successfully applied to quantum annealing schedules, demonstrating its effectiveness in guiding quantum optimizations across various architectures~\cite{Finzgar2024}. In our simulations we model uncorrelated parameter noise, a platform-independent source that captures generic fluctuations. While BO is robust to broader classes of noise, including biased effects, we restrict to evolution times 
$<1 \mu s$, where long-time decay processes are negligible.

Building a statistical model of the landscape to probe \cite{statis} and balancing exploration of unknown regions of the landscape and exploitation of promising known regions, it can provide new guesses of potential optimal set of parameters to try on the QPU.

\section{Running closed loops: numerical simulations and experiments}
\label{sec:results}

\begin{figure*}[!]
    \centering
    \includegraphics[width=\textwidth]{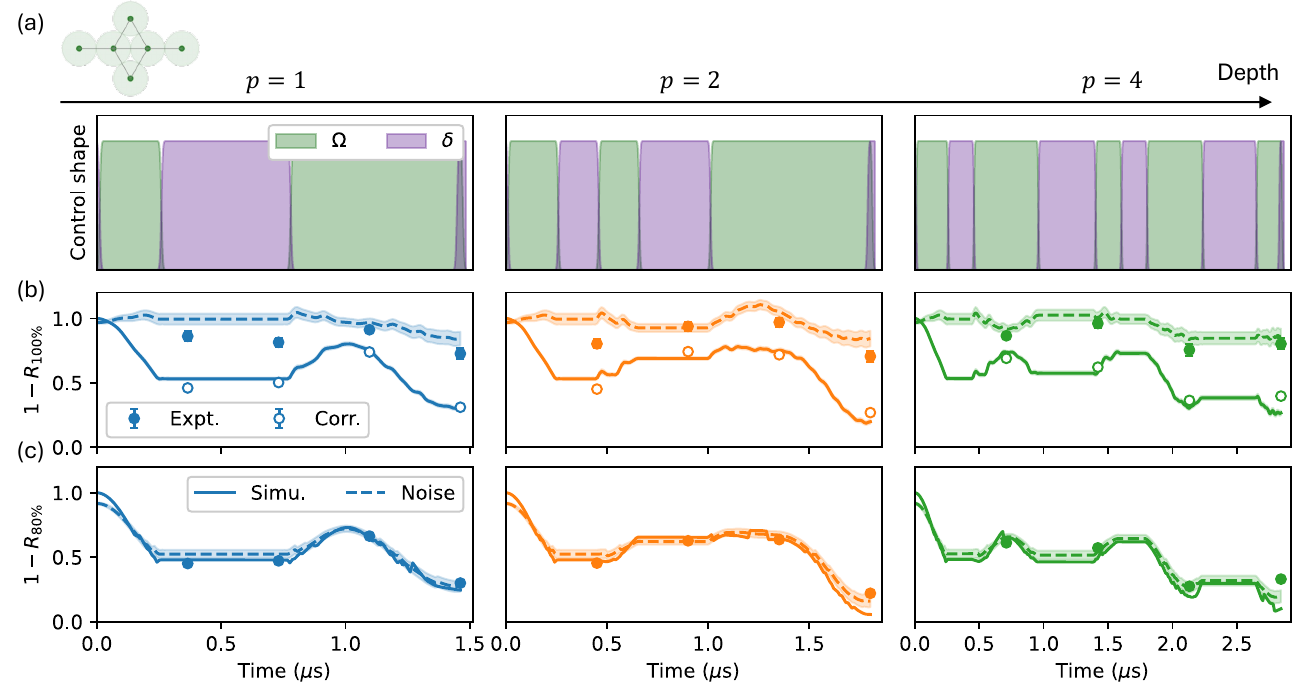}
    \caption{\textbf{Benchmark runs of the quantum dynamics with QAOA-like approach}. (a) Sequences parametrised with QAOA-like approach are applied on the atomic register given in the inset. For increasing depth $p$, the control shapes of $\Omega$ and $\delta$ are displayed as sent to the atoms, i.e. distorted by the shaping device. The evolution of (b) the normalised approximation ratio $1-R_{100\%}$ and (c) the normalised truncated approximation ratio $1-R_{80\%}$ during the dynamics is obtained using noiseless numerical simulations (solid line) or numerical simulations with shot noise ($n_{\rm shots}=1000$) and detection errors ($\varepsilon=3\%$ and $\varepsilon^\prime=8\%$) (dashed). This enables to benchmark raw experimental measurements (filled dots) and SPAM corrected ones (white dots). The standard deviation (filled area-error bars) over the finite sampling with detection errors is obtained using the Jackknife resampling method~\cite{Efron1981}.}
    \label{fig:dyn-qaoa}
\end{figure*}

In this section we present the results of both numerical simulations and runs on the QPU. For the simulations, we constructed graphs with increasing number of qubits ranging from $4$ to $10$, sizes at which simulating all the noises remains tractable. For the experimental part, we selected a $6$-vertex graph from this collection, depicted in Fig.\,\ref{fig:dyn-qaoa} and also introduce two larger graphs, one of size $N=11$ showed in Fig.\,\ref{fig:QAOA} and one of size $N=15$ shown in Fig.\,\ref{fig:graph+finalstate}.

\subsection{Performance with size and measurement precision }
\label{ssec:simul}

To evaluate the performance of our algorithm, we simulate the execution of the Bayesian optimiser using \texttt{Pulser}~\cite{Silverio2022}. This open-source Python package allows to simulate the dynamics of neutral atom systems at the pulse level, while accounting for hardware constraints and noise models specific to Pasqal devices (see App.\,\ref{app:noise}). These features closely replicate the operation of the QPU Orion Alpha used in \ref{ssec:benchmark} and \ref{ssec:exp}.

Fig.~\ref{fig:SimNqubits} presents the results of two sets of experiments performed by: (i) increasing the system size $N \in [4, 10]$ while keeping the number of shots $n_{\rm shots}$ fixed, and (ii) varying $n_{\rm shots} \in \{16, 64, 256, 1028\}$ for a fixed system size $N=6$. 

In the first scenario, performance deteriorates as the number of qubits increases, as shown in the upper panel of Fig.~\ref{fig:SimNqubits}. This trend is expected, as a larger graph leads to a higher probability of measurement errors per qubit. Furthermore, with the number of shots fixed, the reconstructed energy worsens for larger graphs due to the exponential growth in the number of accessible states. This behavior is closely related to the exponential decay in the probability of preparing a maximum independent set, $P_{\rm MIS}$, as reported in Ref.~\cite{Ebadi2022}. In both cases, the scaling originates from the rapid growth of the solution space with graph size, which decreases the probability of optimal solutions and makes them harder to sample reliably.

Although solving the classical problem requires measuring the solution only once, it is crucial to maximise the probability of observing the correct solution, given that it is unknown a priori. Consequently, the improvement in the success rate $S_r$ is more significant than achieving a higher but flat fidelity $R$.

In the second scenario, shown in the lower panel of Fig.~\ref{fig:SimNqubits}, we observe that optimisation is feasible—and sometimes even more effective—using fewer shots. Interestingly, the optimisation with $n_{shots} = 1028$, which we assume to provide a close estimate of the real probability distribution at this size, consistently underperforms compared to those performed with fewer shots. This suggests that reducing the number of shots does not necessarily compromise optimisation results, and may even yield better outcomes in certain cases.

Interestingly, we observe that optimisations with fewer shots may yield better convergence than those with more shots. This can be understood as a trade-off: sampling noise at low shot numbers introduces stochasticity that occasionally aids the optimiser in escaping local optima, whereas higher shot numbers reconstruct a flatter distribution that more faithfully reflects the solution space but can hinder optimisation progress.

\subsection{Benchmarking the dynamics on the hardware}
\label{ssec:benchmark}

We numerically simulate an optimisation procedure for increasing values of $p=1,2,4$ over a parameter space $[0.1\mu$s$, 1\mu$s$]^{2p}$, using $n_{\rm steps}=100$ iterations ($10$ of them being reserved for initialisation) and $n_{\rm shots}=64$ shots per iteration. We obtain three optimised protocols, displayed in Fig.\,\ref{fig:dyn-qaoa}(a), yielding $1-R=0.45,~0.16$ and, $0.27$ respectively. Increasing $p$ from $1$ to $2$ improves the score, but the limited budget of iterations implies that at larger depth, like  with $p=4$, the optimiser will likely not converge, resulting in poorer performance as shown in Fig.\,\ref{fig:dyn-qaoa}(b, c).

We test these three protocols on Orion Alpha, probing the dynamics of $1-R$ at intermediate points in time. This is displayed in Fig.\,\ref{fig:dyn-qaoa}(b), where raw experimental data (filled dots) are compared to noiseless numerical simulations (solid line) and noisy ones (dashed line), the latter being performed with Pulser and including both shot noise and detection errors ($\varepsilon=3\%$ and $\varepsilon^\prime=8\%$). 
In all  cases, $1-R$ gets modified after each layer and reaches a minimum value at the final time $T$. However, a large discrepancy between simulated and experimental data remains. Adding detection errors to the simulation to match the experimental data reveals that the approximation ratio is especially sensitive to $\varepsilon$ due to the asymmetry with $\varepsilon'$ . 
As more population is transferred to MIS states, the number of measurements violating the Rydberg blockade constraint is increasing as non-zero $\varepsilon$ produces non IS bitstrings. With detection errors included, the final values of $1-R$ remain very close to $\approx 1$, i.e. its initial value, making the idea of directly optimising on the QPU challenging. In the following section and in Appendix~\ref{app:spam} we use mitigation protocols to keep this effect under control.

\subsection{Correcting for detection errors}
\label{ssec:errors}

As explained above, the approximation ratio is particularly sensitive to SPAM errors, while decoherence effects at the durations considered here ($T<2.5 \mu s$) are essentially negligible compared to such large SPAM errors. Since including laser phase noise in the simulations is quite resource intensive, as it requires sampling from a full power density spectrum that must itself be experimentally characterized on the QPU. Thus we have considered to correcting only for the detection errors. 

A first solution in this direction is the scheme described in App.\ref{app:spam}, at the expense of possible additional computing costs. With $\varepsilon=3\%$ and $\varepsilon'=8\%$, the corrected data, shown with empty dots in  Fig.\,\ref{fig:dyn-qaoa}(b), are more in line with noiseless simulation, and the remaining gap between the two at longer times can be attributed to decoherence effects. It is worth noting that imperfect correction of the measured distributions can nonetheless happen due to finite sampling effects and wrong estimation of $\varepsilon$ and $\varepsilon^\prime$. 

A second possibility to make the approximation ratio more resilient to detection errors is to discard all the bitstrings violating the Rydberg blockade condition, assuming that their presence can only be due to noise. In the hard blockade limit where $U\gg\delta,\Omega$, the resulting discarding slightly modify the initial distribution and is a possible way to mitigate $\varepsilon$ since it erases its impact. However, in the regime of control parameters of this implementation, we can not rule out that the dynamics itself, through facilitation mechanisms \cite{Marcuzzi2017-wh} for instance, could not produce bitstrings with unwanted excitations. Discarding bitstrings with the previous rule may alter significantly the optimisation landscape, as some measured distributions could be almost emptied. Another straightforward approach consists in erasing a fixed percentage of each measured distribution, discarding bitstrings $\boldsymbol{z}$ among the ones with the highest energies $C_\Graph(\boldsymbol{z})$. This percentage can either be estimated knowing the value of $\varepsilon$ or set arbitrarily high, at the expense of maybe discarding bitstrings not produced by the imperfect detection events. As $N$ increases, the effect of detection errors intensifies and a large discarding percentage might be needed, at the expense of acquiring more statistics to balance the loss. 

The effect of this truncation is shown in Fig.\,\ref{fig:dyn-qaoa}(c) with $R_{80\%}$ being the approximation ratio computed when keeping only the $80\%$ best bitstrings of the distributions.  These percentage has been chosen in order to balance data retention and error suppression. Both numerically simulated noisy and experimental refined data present good agreement with numerically simulated noiseless ones when discarding $20\%$ at $N=6$. 
Overall, at this timescale, the main effect impacting the MIS preparation remains the detection errors and possible miscalibration of the controls, as emphasised in the next part. However, by optimising directly on the QPU, one can still locate an optimum as long as the landscape is not too flattened.

\subsection{Experimental closed loops}
\label{ssec:exp}

\begin{table}[b]
\centering
\begin{tabular}{|c|c|c||c|c|c|}
\hline
N  & $n_{\rm steps}$ & $n_{\rm shots}$ per step & $R$   & $F$   & $S_r$ \\ \hline\hline
6  & 110             & 32                       & 0.817 & 0.438 & 7     \\ \hline
6  & 110             & 64                       & 0.808 & 0.438 & 7     \\ \hline
11 & 110             & 32                       & 0.646 & .094  & 1.667 \\ \hline
\end{tabular}
\caption{Parameters of the loops performed on the machine and final approximation ratio $R$, fidelity $F$ and solution ratio $S_r$.}
\label{tab:loopsRun1Shots}
\end{table}

We perform closed-loop optimisation directly on the QPU while implementing a $20\%$ discard bitstring method at each step. The number of shots used per step is kept deliberately low for two key reasons: first, to efficiently manage the fixed budget of computational time, and second, to evaluate the effectiveness of BO under such conditions. The specific properties of each performed loop and the results are summed up in Table~\ref{tab:loopsRun1Shots}. For the $N=6$ qubits graph the results are satisfying for approximation ratio $R$, fidelity $F$ and Solution ratio $S_r$. Although optimisation is rather successful also for $N=11$ (final $S_r$ tells us that the solution state stands out), the $R$ and $F$ results show that for this instance, more shots seem to be needed for a more reliable optimisation.


\begin{figure}[!]
    \centering
    \includegraphics[width = \linewidth]{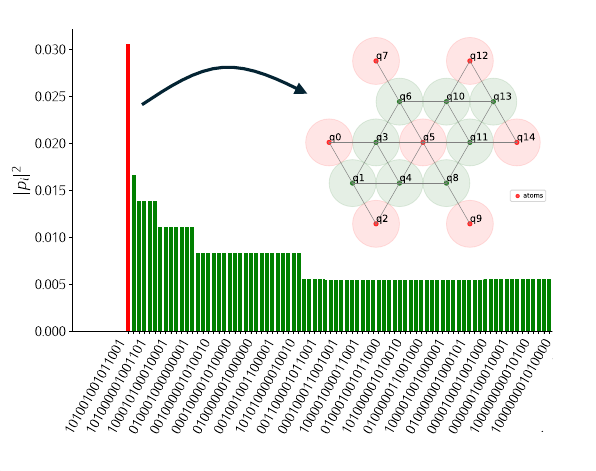}
    \caption{\textbf{MIS solution state obtained for the $N=15$ graph}. The largest graph of $N=15$ qubits used for the experiment is plotted as an atomic register, with red qubits highlighting the MIS solution state we aim to find, corresponding to the target bitstring. The probability distribution obtained by sampling the final state produced using BO on QAOA is represented as a histogram. For each of the $2^{15}$ possible bitstrings, their probabilities $p_i$ were calculated as the ratio of the number of times they were measured to the total number of shots. The long tail of the histogram corresponds to states that were measured only once.
}
    \label{fig:graph+finalstate}
\end{figure}

\begin{figure*}[!]
    \centering
    \includegraphics[width = \linewidth]{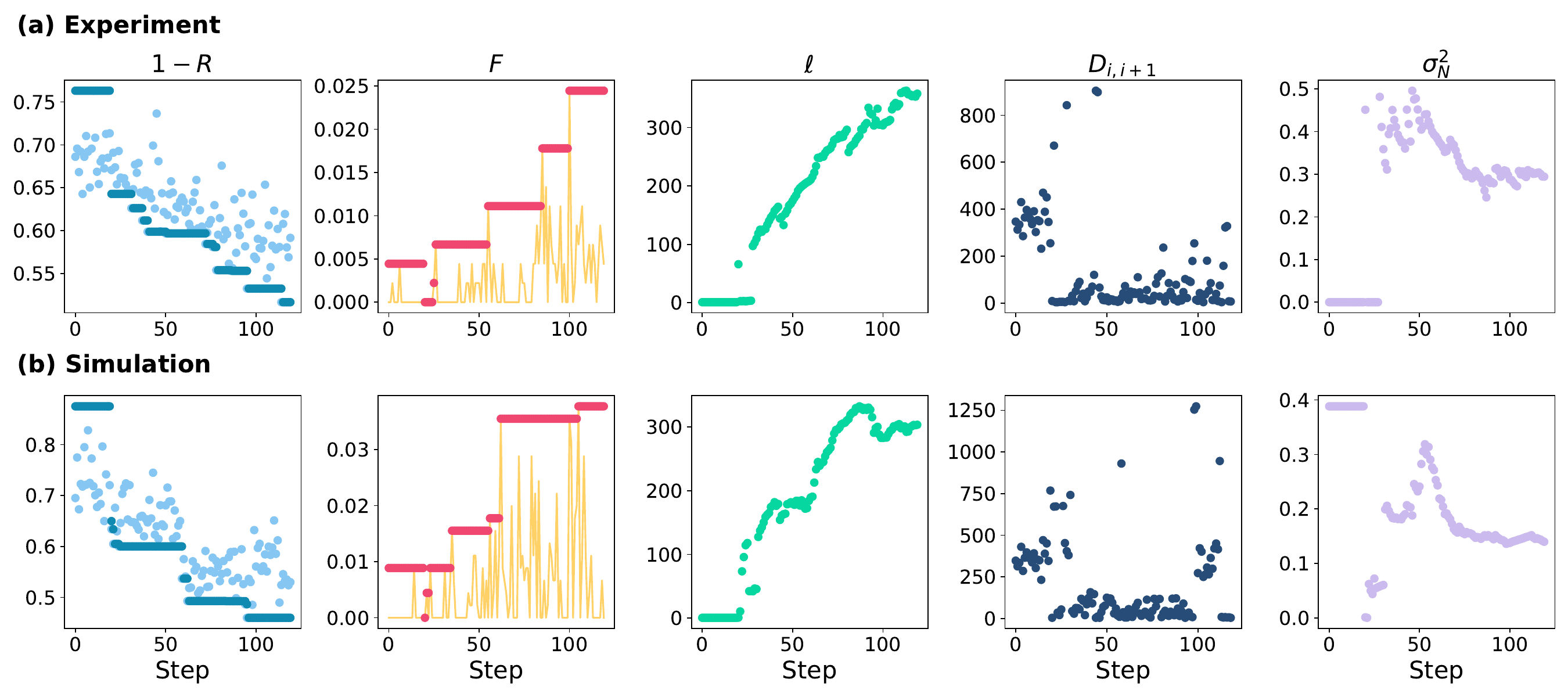}
    \caption{\textbf{Evolution of key metrics through the closed-loop optimisation on the $15-$qubit graph}. 
    (a) Experimental and (b) numerically simulated results compared. For each one we show during optimisation the following parameters (from left to right): the complement of the success rate $(1-R)$, fidelity $F$, correlation length $\ell$, distance between consecutive iterations $D_{i, i+1}$, and Gaussian Process Noise $\sigma$. Experimental data demonstrate gradual convergence and variations due to noise, while numerical simulations provide a controlled reference for comparison. Solid lines and markers indicate trends and discrete updates, respectively. Highlighted plateaus correspond to regions of stability in the optimisation dynamics.}
    \label{fig:run3loops}
\end{figure*}

Finally, we run one optimisation loop on a $N=15$ qubits graph (inset of Fig.\,\ref{fig:graph+finalstate} with the MIS solution highlighted in red) with depth $p=5$. 

The experimental optimisation was successful: in Fig.\,\ref{fig:graph+finalstate} we show the final state with the solution bitstring measured more often than all the other ones. The loop was run using all the remaining available shots which were in the order of $50 000$. We decided to run $100$ steps of optimisation with 20 initial points of training at $450$ shots per step to gain the best reconstruction of the state possible, as mentioned in the previous subsection. Considering the discard of $20\%$ we are optimising with 360 effective shots.

In Figure~\ref{fig:run3loops} we plot the QAOA parameters obtained at each step of optimisation, both for the simulated and the experimental cases on the $N=15$ graph. They show similar trends and can therefore be commented together. Clearly, $R$ and $F$ increase during optimisation showing that the algorithm is converging to a solution. 
Along these parameters, listed in the previous subsection, we analyse the BO parameters. 
These are (see Eq.~\eqref{eq:maternKernel}):
\begin{itemize}
    \item $\ell$: the correlation length, which is a parameter that controls to which distance points are considered correlated by the Gaussian process,
    \item $\sigma_N^2$: a noise parameter added to the kernel's diagonal to account for noise in the energy of the unknown function and learnt at each step to fit noise,
    \item $D_{i, i+1}$: distance in the parameters' space between two consecutive sampled points, to check if the algorithm is under exploration of the landscape or exploitation (sampling points further or closer to each other).
\end{itemize}

The BO parameters during the closed loop on the big instance show a well performed optimisation loop. The correlation length keeps rising, meaning that BO keeps exploring the landscape for better points at each step, without necessarily stopping on one. This is seen also on the distance $D_{i, i+1}$ that alternates between larger values or jumps (in the order of the hundreds) in the parameters space with few local jumps (in the order of the tens). This pattern is typical of BO that concentrates on an area and if not satisfied jumps to another part of the landscape.

The main difference between simulated an experimental data is the noise level $\sigma^2_N$, which is a parameter that is learnt by the algorithm at each optimisation step to overcome the difference in the energy values of the landscape that it predicts compared to the real ones obtained. This parameter is initially large and then settles around a specific value which is larger in the experimental scenario (almost double) compared to the numerical simulation. This means that even if we accounted for noise in the simulation, the real source of noise in the QPU was larger and thus affected more strongly the results. 

We can say with confidence that we would have got even better results with more steps of optimisation, this is what the trend in the parameters clearly indicates. However, despite the fact that we reach a relatively low fidelity (about $3\%$), a solution ratio of 1.8 was reached, which is not trivial given the dimension of the graph and the limited number of runs available.

\section*{Conclusion and outlook}
\label{sec:conclusions}
In this work, we have demonstrated the feasibility and advantages of employing analog QAOA with BO on a neutral atom QPU. By focusing on the MIS problem, we showcased how the physical properties of the Rydberg atom platform could be exploited for efficient problem encoding and solution retrieval. The integration of BO proved crucial in efficiently navigating the parameter space under stringent resource constraints, achieving significant convergence with limited computational overhead.

Experimental results corroborated numerical simulation predictions, highlighting the robustness of the approach despite hardware-induced noise and imperfections. The study also underscored the importance of tailored noise mitigation strategies, such as selective bitstring discarding and error correction, to improve algorithmic fidelity and accuracy.

This work establishes a framework for scaling analog quantum algorithms to larger problem instances without having to rely on an increasing number of measurements. When considering larger graphs, an identified limitation of the QAOA approach is the need for increased $p$ to further improve a score, leading to longer sequences, that allow for more quantum error, and to longer convergence time, as the parameter space expands. One interesting idea to circumvent this issue consists in making an educated guess from previous layers to the new one~\cite{Zhou20}, hence limiting the size of the parameter space to optimise on. 

Thus, from the algorithmic side, more work is needed to confirm the efficiency of the method to larger systems and more complex problem classes, while, from the experimental side, future directions of studying  should include more advanced noise mitigation techniques to improve performance on NISQ devices.

\backmatter

\bmhead{Acknowledgements}

\section*{Declarations}
\bmhead{Funding}
The research of S.T., E.T. and E.E. is partially funded by the International Foundation Big Data and Artificial Intelligence for Human Development (IFAB, project “Quantum Computing for Applications”). E.E. also acknowledges financial support from the National Centre for HPC, Big Data, and Quantum Computing (Spoke 10,
CN00000013) and the 2022-PRIN Project “Hybrid algorithms for quantum simulators”.
The experimental runs on the Pasqal device have been made possible via the CINECA ISCRA-C project "BATQUO".
\bmhead{Conflict of interest/Competing interests}
Not applicable
\bmhead{Ethics approval and consent to participate}
Not applicable
\bmhead{Consent for publication}
Not applicable
\bmhead{Data availability}
Data supporting the results in this work can be obtained from the corresponding author upon reasonable request.
\bmhead{Materials availability}
Not applicable
\bmhead{Code availability}
Code supporting the results in this work can be obtained from the corresponding author upon reasonable request.
\bmhead{Author contribution}
Not applicable

\begin{appendices}

\section{Analog computation with neutral atoms}
\label{app:Analog}
\subsection{Ising model with ground-Rydberg encoding}
The experimental setup consists of spatial configurations of $^{87}$Rb atoms trapped in an optical tweezer array. A Spatial Light Modulator (SLM) can imprint various phase patterns on the trapping beam, generating arbitrary spatial register in 2D, and even 3D \cite{Barredo2018}. Cooled atoms half-fill the traps and a moving tweezer, combined with a classical rearrangement algorithm, allow assembling $N$ atoms at specific positions with efficiency $p_{\rm move}^N$. A fluorescence image can be acquired, revealing with bright spots the filled traps and thus the atomic presence. Those atoms can be excited from their ground states to Rydberg states, highly-excited electronic states of the atoms energy spectrum, to make them interact with dipole-dipole interactions. Choosing a specific Rydberg state in the excited manifold enables to engineer the type and strength of the interactions. \\

In this work, we encode the qubit state $0$ in the atom ground state $\ket{0}=|5S_{1/2},F=2,m_F=2\rangle$ and $1$ in a specific Rydberg state $\ket{1}=|60S_{1/2},m_J=1/2\rangle$, naturally realising for $N$ atoms, the global Ising Hamiltonian 
\begin{equation}
    \hat{H}_{\boldsymbol{\theta}}(t)/\hbar = \frac{\Omega_{\boldsymbol{\theta}}(t)}{2}\sum_{i=1}^N \hat{\sigma}_i^x - \delta_{\boldsymbol{\theta}}(t) \sum_{i=1}^N \hat{n}_i + \sum_{i \neq j} U_{ij} \hat{n}_i \hat{n}_j
    \label{eq:ising-hamiltonian}
\end{equation}
where $\hat{\sigma}_i^\alpha$ are Pauli matrices, $\hat{n}_i=(\hat{\mathbbm{1}}+\hat{\sigma}_i^z)/2$, the Rydberg occupation number at site $i$. The transition $\ket{0}\leftrightarrow\ket{1}$ is driven using a two-photon scheme with two lasers with wavelengths $1013/420$nm. This coupling can be effectively modelled by two time-dependent and parameterised control fields : $\Omega_{\boldsymbol{\theta}}(t)$, the Rabi frequency/amplitude of the drive and $\delta_{\boldsymbol{\theta}}(t)$ the detuning from the resonance of the used Rydberg transition. $\boldsymbol{\theta}$ is a set of parameters which usually defines how the controls are shaped in time. Finally, the pairwise van der Waals (vdW) term reads $U_{ij}=C_6/\hbar/r_{ij}^6$ with $C_6$ a constant depending on the chosen Rydberg state and $r_{ij}$, the interatomic distance between atoms $i$ and $j$. Here, $C_6/\hbar\approx 2\pi\times 137$ GHz$\cdot\mu$m$^6$ and $r_{ij}>5~\mu$m due to trapping constraints. The measurement process enables to map the qubit logical state to the occupancy of the trap containing the atom, revealing thus a bright/black spot when an atom is measured in $\ket{0}/\ket{1}$.\\

The vdW interactions in Eq.\ref{eq:ising-hamiltonian} actually engineer the \textit{Rydberg blockade} effect, shifting the energy of doubly excited states $\ket{1}_i\ket{1}_j$. The simultaneous excitation with amplitude $\Omega$ of pairs of atoms closer than the blockade radius $r_b = (C_6/\hbar \Omega)^{1/6}$ is prevented. In the meantime, interactions terms for atoms separated by far more than $r_b$ can be neglected due to the sharp decay of $U_{ij}$ with the distance. For a blockaded pair initialised in $\ket{00}$, resonantly  addressing the $\ket{0}\leftrightarrow\ket{1}$ transition (i.e. with $\delta=0$) drives the system to the $(\ket{01}+\ket{10})/\sqrt{2}$ with a coupling enhanced to $\sqrt{2}\Omega$, instead of driving it to $\ket{11}$. This blockade effect is the stepping stone used for embedding and solving combinatorial problems in neutral atom platforms. \\

\subsubsection{Analog evolution and computation cycle}
The ability to evolve a quantum system in time starting from a simple initial state is at the heart of the quantum computation paradigm. The quantum state obtained when globally evolving an atomic system under the above Hamiltonian for a parameterised duration $T_{\boldsymbol{\theta}}$ reads 
\begin{equation}
   \ket{\psi(\boldsymbol{\theta})}=\mathcal{T}\left[\exp(-\frac{i}{\hbar}\int_{s=0}^{T_{\boldsymbol{\theta}}}\hat{H}_{\boldsymbol{\theta}}(s)ds)\right]\ket{0}^{\otimes N}
   \label{eq:evolution}
\end{equation}
with $\mathcal{T}$ the time-ordering operator. In the analog mode, the global Hamiltonian, obtained in the ground-Rydberg encoding, is directly fine-tuned by playing with either the duration, shape, amplitude and frequency of the controls $\Omega_{\boldsymbol{\theta}}(t),\delta_{\boldsymbol{\theta}}(t)$ or the atomic positions $\boldsymbol{r}_i$: it constitutes a resource for computation. With the ground-ground qubit encoding, neutral atom technology can tackle the digital mode which relies on a discrete set of quantum operations, or $\textit{gates}$, instead of a continuous evolution.\\

While the actual quantum dynamics happens at the $\Omega/2\pi\sim$ MHz scale, the necessity to load the tweezers ($\sim 100$ms), arrange the atoms ($\sim 150$ms), initialise their state ($\sim 50$ms) and image them several times ($\sim 100$ms) for each shot reduces the repetition rate of the useful computation to a few Hz. In addition, the effective repetition rate is even scaled down by the probability of assembling a defect-free quantum register at each cycle. Note that in practice, to obtain $n_{\rm shots}$ calculations with a properly arranged register, we ask for $n_{\rm shots}\times p_{\rm move}^{-N}$ in the following.

\section{Description of the Bayesian Optimisation method}
\label{app:Bayes}
We describe here briefly the Bayesian Optimisation (BO) algorithm (for a review see~\cite{Snoek2012,Shahriari2016}). In~\cite{TibaldiBO2023}, we studied its behavior compared to other
global optimizers, showing that the convergence rate to a local minimum is faster. We also demonstrated that the Bayesian approach is efficient in terms of the number of circuit runs, while being robust against Gaussian noise sources. These characteristics make BO a promising tool for implementing QAOA efficiently on a noisy intermediate-scale quantum device.

This algorithm finds the extremum of an unknown (target) function $f(\theta)$ that depends on a set of parameters $\theta$ with few evaluations of the function.
This algorithm exploits two subroutines: a Gaussian Process and an Acquisition Function. The former is a probability distribution over the unknown parameters that reconstructs the landscape of the target function and provides an uncertainty on such reconstruction. The latter is a function of the parameter space that combines the prediction and uncertainty of the Gaussian Process and provides the likelihood of finding an extremum point. 

\subsection{Gaussian Process} 
\begin{figure}
    \centering
    \includegraphics[width = \linewidth]{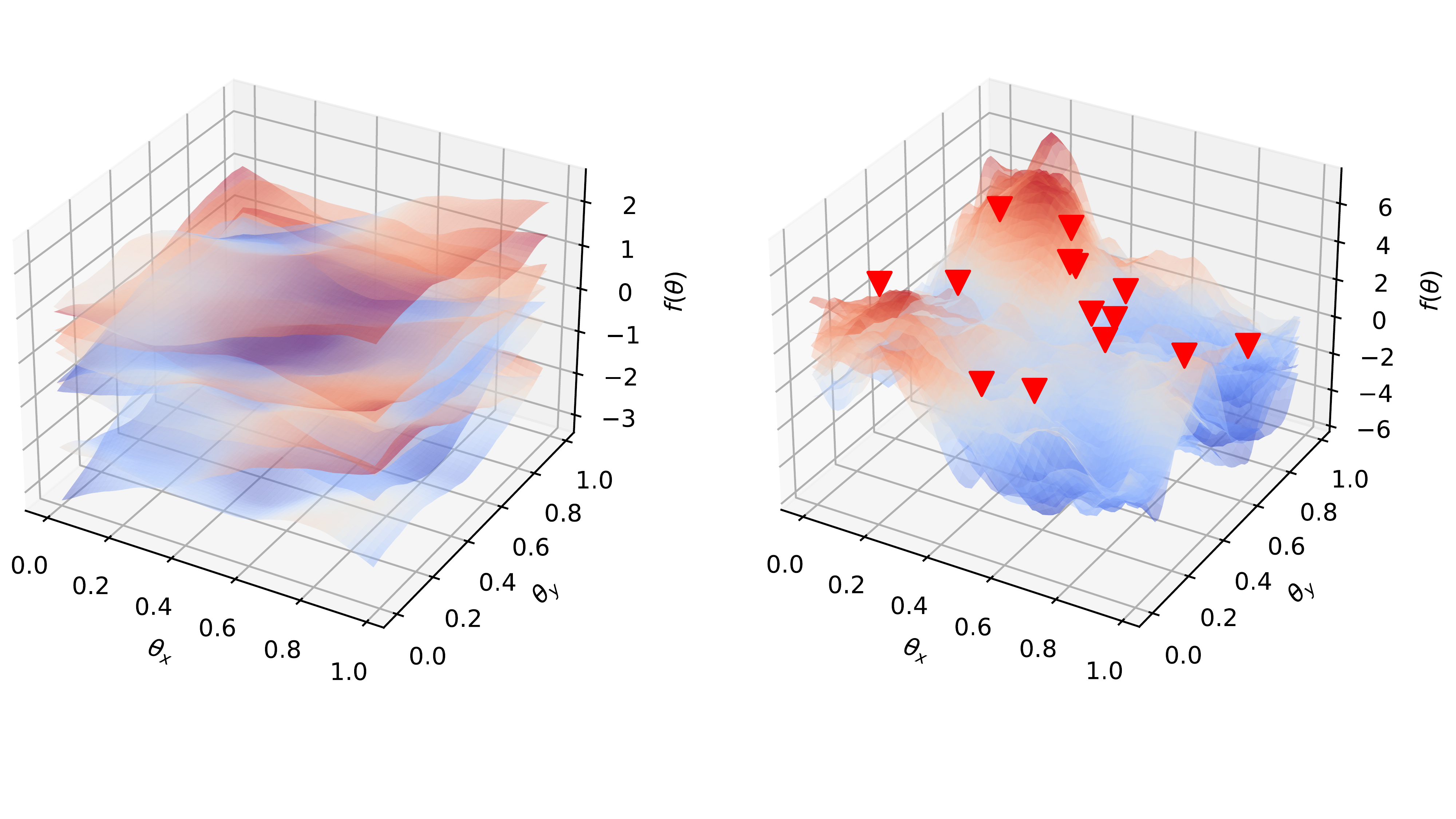}
    \caption{(left) Functions sampled from the prior of the Gaussian Process with $\ell = 1, \sigma^2 = 1$. For multiple points of the parameter space we sample from the Gaussian prior and obtain smooth manifolds with values averaged to 0. (right) Functions sampled from the Gaussian posterior after sampling the function $f(\theta)$ at the red arrow points. Now the sampled manifolds pass from the same points, while have different values away from them, that is the variance of the posterior.}
    \label{fig:prior-posterior}
\end{figure}

A Gaussian Process (GP) is a multinomial distribution created from a mean function $\mu$ and a correlation function $k$:
\begin{equation}
\label{eq:multinomial}
    \mathcal{N}(\mu(\theta),k(\theta_i, \theta_j)).
\end{equation}
Both $\mu$ and $k$ depend on the parameters $\theta \in \mathbb{R}^N$ . When we sample~\eqref{eq:multinomial} over the parameter space we get a set of points distributed with mean value $\mu(\theta)$ and variance $k(\theta_i, \theta_i)$ and correlated with each other according to the correlation matrix with entries given by $k(\theta_i, \theta_j)$, an example is shown in Fig.\,\ref{fig:prior-posterior}. The purpose of the GP is to reproduce faithfully the output of the unknown function. To do this, we have to make some initial assumptions. For example, we can set, $\mu=0$ which means that we assume our function has values distributed around 0. And we can make the points being correlated according to a smooth function like the Matérn function:
\begin{equation}
\label{eq:maternKernel}
    k(\theta_i, \theta_j) = \sigma^2 \left(1 + \frac{\sqrt{3}||\theta_i - \theta_j||}{\ell} \right) e^{-\frac{\sqrt{3}||\theta_i - \theta_j||}{\ell}} + \sigma^2_N
\end{equation}
This kernel function depends on three \textit{hyperparameters} $\sigma^2, \ell$, $\sigma^2_N$. The first one is interpreted as a normalisation term. The second one is the \textit{correlation length}, an important element which tell us at which distance we can consider the points correlated. The last one accounts for noisy estimations of the landscape energy. By setting $\mu$ and $k$ we have created a (very general) \textit{prior} $p(f)$ for the Gaussian process which generates smooth functions with values centered around 0 and variance, or uncertainty, $k(\theta_i, \theta_i) = \sigma^2$.

If we want the GP to produce functions that resemble the target $f$ we need to update it with data, that is a collection of $M$ sampled points \{$\blambda, \by\} = \{\lambda_i, f(\lambda_i)\}_{i=1}^M$. This is done mathematically by the  \textit{conditioning} operation which produces a new multinomial $\mathcal{N}'$, the \textit{posterior} $p(f|\by)$, with new mean $\mu'$ and variance $k'$ functions:
\begin{eqnarray}
    \label{eq:posterior}
    \mu'(\theta) &=& \bkappa^T \cdot K^{-1} \cdot \by \\
    k'(\theta) &=& k(\theta, \theta) - \bkappa^T \cdot K \cdot \bkappa
\end{eqnarray}
with $\bkappa = k(\theta, \blambda)$ is the array obtained calculating the correlation between the generic point $\theta$ and each data point $\lambda_i$, while $K$ is the $M \times M$ correlation matrix of the data with entries $K_{i,j} = k(\lambda_i, \lambda_j)$. Now, if we sample the posterior $\mathcal{N}'$ many times over the parameter space each generated function calculated at $\lambda_i$ will produce exactly $f(\lambda_i)$ with variance 0, it will give values closes to $f(\lambda_i)$ in the neighbourhood of $\lambda_i$ with variance increasing with the distance and then the points will again be randomly distributed with large variance away from the sampled points $\blambda$, visual example in Fig.\,\ref{fig:prior-posterior}.

The common practice to choose the hyperparameters of the kernel function $\sigma^2, \ell, \sigma^2_N$ is to maximise the log-marginal likelihood $p(\by | f)$ of the posterior~\cite{Rasmussen2005}, which is the integral over all possible functions of the prior $p(f)$ times the likelihood of the model. To sum up, every time we give a set of sampled points to the GP we perform the fitting operation which requires, maximising the likelihood to obtain the hyperparameters of the kernel and then calculating the new mean $\mu'$ and variance $k'$~\eqref{eq:posterior} which can be used to make updated predictions on the landscape of the target function $f$.

\subsection{Acquisition Function}
After fitting the GP we have a current best value $f_m$ which is the lowest point found up to now. We now need a function of the parameter space that returns the likelihood of finding a point lower than $f_m$. A common choice for this is the Expected Improvement (EI), defined as:
\begin{equation}
\label{eq:EI}
    \text{EI}(\theta) = \Phi(z)(f_m - \mu') + \phi(z)k',
\end{equation}
where $\mu', k'$ are the posterior mean and variance of Eq.~\eqref{eq:posterior}, while $\Phi(z)$, $\phi(z)$ are the cumulative distribution function and the probability density function of the posterior, calculated with the rescaled variable $z = (f_m - \mu')/k'$. Recall that $\mu', k'$ depend on the position $\theta$ of the parameter space where we sample, so EI depends only on the parameters $\theta$. Intuitively, the two terms of~\eqref{eq:EI} represent the exploration-exploitation trade-off of global optimisation: the first term favours the search for a new minimum (being dependent on the difference between the posterior mean $\mu'$ and the current best point $f_m$) whereas the second one favours the research in parts of the landscape with the largest variance, that is less explored.

\section{Hardware noise sources}
\label{app:noise}
Several experimental imperfections can alter the quantum computation phase, with hardware-related effects impacting at a higher level the efficiency of algorithms. The most detrimental sources of noise are listed in the following, and their effects can be simulated using the open-source package, \texttt{Pulser}~\cite{Silverio2022}. 
Errors can first occur at the initialisation stage, where qubits are prepared in $\ket{0}$ through optical pumping with efficiency $\eta$. Imperfect initialisation can not be easily detected, and the atom is then always measured in $\ket{0}$, remaining unaffected by controls or interactions during the dynamics. A typical value for the $\textit{state preparation}$ error per qubit is $\eta\approx0.5\%$. \\
During the quantum dynamics happening in Eq.\ref{eq:evolution}, the traps are turned off and the atoms are freely moving due to their residual temperature $T\sim10~\mu$K. This random thermal motion affects the frequency of the light the atoms "see" and results in \textit{Doppler effect}, randomly shifting the detuning $\delta$ for each qubit. This noise can thus be modelled by additional terms $\delta_i$ generated from a centred normal distribution with deviation $\sigma_\delta(T)=||\vec{k}||\sqrt{k_BT/m}$, with $||\vec{k}||\approx 2\pi\times 1.38\mathrm{\mu m}^{-1}$ the norm of the wave vector of the effective laser system, $k_B$, the Boltzmann constant, and $m$, the $^{87}$Rb mass. Thus at $T=50~\mu$K, $\sigma_\delta(T)/2\pi\approx0.1$ MHz to compare with used value of $\Omega,\delta$. \\
Fluctuations in the beam-delivery system, such as optical fibre, can also alter the value of the Rabi frequency from one shot to another. Effectively, $\Omega$ will follow a Gaussian distribution centred around its calibrated value and spread with standard deviation $\sigma_\Omega$. Moreover, excitation lasers generally exhibit a Gaussian intensity profile resulting in a distance dependent damping following $\Omega(r)=\Omega_{max}\exp[-(r/w)^2]$ with $w\approx 180~\mu$m, the laser waist and $r$ distance from the laser focus point on the array. For the relatively small register instances of this work, this last effect remains negligible. \\
Additional effects, such as spontaneous emission or laser phase noise, can induce decay or dephasing during the dynamics. These processes can be modelled using effective noise channels, characterized by the timescales $T_1\sim100~\mu$s and $T_2\sim5~\mu$s, for the decay and dephasing channels respectively. As the experimental protocols tested on the QPU and presented in this paper last no longer than $2.8~\mu$s, we neglect the influence of those noise channels in our simulations. \\

\section{Correcting for detection errors during the imaging process}
\label{app:spam}
The measurement process is inherently flawed by several physical processes which can result in measuring a $1$ instead of a $0$, leading to a \textit{false positive} detection event, and conversely, to a \textit{false negative}. Background-gas collisions can eject a recaptured atom, emptying a trap and thus flipping a $0$ to a $1$ in a bitstring with probability $\varepsilon \approx 1\text{--}3\%$. Additionally, the ejection of atoms in Rydberg states typically lasts around a few microseconds, leaving enough time for some of them to decay from $\ket{1}$, effectively flipping a $1$ to a $0$ in a bitstring with probability $\varepsilon^\prime \approx 5\text{--}8\%$. Reducing the temperature of the background atoms helps lower $\varepsilon$, while increasing the Rydberg lifetime by using larger Rydberg levels can lower $\varepsilon^\prime$. Let us remark that such relatively large range of values stems from independent calibration measurements of the detection process, which are subject to statistical fluctuations and experimental drifts over days/weeks. The value of $\varepsilon'$ is calibrated before an experiment and does not change drastically if the data are taken within a few days. 

The many physical processes inducing bit flips during the measurement phase can be effectively encompassed by two terms, $\varepsilon = p(0 \rightarrow 1)$ and $\varepsilon^\prime = p(1 \rightarrow 0)$. Keeping in mind that this definition depends on the physical states chosen as $\ket{0}$ and $\ket{1}$, the values of these two terms can vary for different experiments but are usually at the percent level. Moreover, when the atoms are addressed locally, these values can become site-specific. Modeling these bit flips can be achieved using the following transfer matrix:
\begin{equation}
    M_i = \begin{pmatrix}
        1 - \varepsilon_i & \varepsilon_i^\prime \\
        \varepsilon_i & 1 - \varepsilon_i^\prime
    \end{pmatrix},
    \label{eq:transfer-matrix}
\end{equation}
The incorrectly measured distribution is thus $\tilde{P}_\psi = (\bigotimes_i M_i) P_\psi$, where we assume uncorrelated errors. While the detection errors at the single-qubit level remain low and easy to correct, they quickly scale with the size of the system. For instance, for $N = 100$ and $\varepsilon = 1\%$, measuring $\ket{\psi} = \ket{0}^{\otimes N}$ is only achieved with an efficiency of $(1 - \varepsilon)^N = 36.6\%$.

Correcting these errors becomes critical for state preparation or algorithmic tasks and requires the inversion of a $2^N \times 2^N$ matrix. While the matrix construction/inversion procedure can be sped up using tensor formalism, the most computationally resource-demanding aspect lies in building the probability distribution vector of size $2^N$. Moreover, due to finite sampling of the state and incorrect estimation of $\varepsilon, \varepsilon^\prime$, $(\bigotimes_i M_i^{-1}) \tilde{P}_\psi$ may not be a proper probability distribution. While naive methods such as renormalization or truncation can provide sufficient approximations of $P_\psi$, more advanced methods such as Bayesian reconstruction may yield more accurate results. However, such a method is currently effectively limited to $N = 25$ to remain within a few minutes of computation.

\end{appendices}

\bibliography{biblio}

\end{document}